\documentstyle[prl,aps]{revtex}
\def\mathrm#1{{\rm #1}}
\def\PR{{\it Phys. Rev.}}
\def\PRL{{\it Phys. Rev. Lett.}}

\def\sqr#1#2{{\vcenter{\vbox{\hrule height.#2pt
                       \hbox{\vrule width.#2pt height#1pt \kern#1pt
                       \vrule width.#2pt}
                     \hrule height.#2pt}}}}
\def\square{\mathchoice\sqr34\sqr34\sqr{2.1}3\sqr{1.5}3}

\begin{document}

\twocolumn

\title{Universal conductance distribution
in three dimensional systems in high magnetic fields}

\author{Tomi Ohtsuki, Keith Slevin\dag and
Tohru Kawarabayashi\ddag}

\address{Department of Physics, Sophia University, Kioi-cho 7-1, Chiyoda-ku,
Tokyo 102-8554, Japan}

\address{\dag Department of Physics, Osaka University,
Machikaneyama 1-16, Toyonaka 560, Japan
}

\address{\ddag Institute for Solid State Physics, University of Tokyo,
Roppongi 7-22-1,
Minato-ku, Tokyo 106-8666, Japan}

\maketitle

\begin{abstract}
The nature of the critical point of the Anderson transition
in high magnetic fields is discussed with an emphasis
on scale invariance and universality of the critical exponent.
Special attention is paid to the distribution function of the
conductance which becomes size and model independent at the critical point.
The fractal properties of the wave function which are related to 
scale invariance are also discussed.
\end{abstract}

\section{Introduction}
The Anderson transition (AT) has been attracting the attention of
condensed matter physicists for more than four decades \cite{Anderson,Mott}.
The AT is a zero temperature quantum phase transition separating
metallic and insulating phases which is
induced by a spatially fluctuating random potential.
The transition can be described using the scaling theory of localization \cite{gang,kawabata}.
Near the critical point in three dimensional (3D)
systems behavior which is typical of quantum phase transitions is
observed for quantities such as the conductance $g$, correlation
length $\xi$ etc. 
For example, as we approach the critical point by changing a parameter
$w$ such as strength of disorder or Fermi energy,
the correlation length diverges as
\begin{equation}
\xi\sim |w-w_c|^{-\nu} ,
\end{equation}
while the conductivity $\sigma$ vanishes from the metallic side
according to the power law,
\begin{equation}
\sigma\sim |w-w_c|^s .
\label{sexp}
\end{equation}
If we approach the transition from the insulating side, then the dielectric constant
$\epsilon$ diverges as
\begin{equation}
\epsilon \sim |w-w_c|^{-s'} .
\end{equation}

As in the critical phenomena of magnetic systems, the exponents $\nu$, $s$
and $s'$
are not independent but are related \cite{wegner}
\begin{equation}
s = (d-2)\nu ,{\quad } s'=2\nu .
\label{eqn:wegner}
\end{equation}
A knowledge of $\nu$ is enough to fix the critical exponents $s$
and $s'$.
These latter exponents can be measured experimentally though there has
been controversy concerning the correct values \cite{Itoh}. It
is thus important to have a precise theoretical estimated of 
$\nu$ in order to compare with the experiments.

The value of $\nu$ is expected to be universal, i.e., independent of 
the details of the model and dependent only on 
basic symmetries such as that under the operation of time reversal.
The classification of the critical behavior
according to the symmetries of the system
was predicted from field theoretic
considerations \cite{field,Hikami}, and recently verified numerically \cite{SO}.

At the Anderson transition the correlation length diverges and the
wave function becomes scale invariant.
This invariance is characterized by a fractal dimension $D_2$.
This is reflected in the size independence of the distribution function
of the conductance \cite{SO,markos} as well as the statistics of
the energy levels \cite{Shk,ZK,OO,BSZK,KOSO}.

In this paper we report a numerical simulation of the wave function
dynamics at the critical point for the tight binding model 
in a magnetic field.
We have observed anomalous diffusion and estimated
the fractal dimensionality.
The universality of the distribution function of the conductance at 
the critical point has also been verified.
The final section is devoted to summary of our results
and concluding remarks in connection with experiments.

\section{Anderson transition in magnetic fields}
%
%
Magnetic fields have two effects on the Anderson transition.
One is to delocalize the electronic states by breaking
time reversal symmetry,
and the other is to localize the electronic states by
shrinking the wave function due to cyclotron motion.
Which of these two dominates depends on the situation.

The tight binding Hamiltonian which incorporates
the effect of magnetic fields is given by
\begin{equation}
 H = V \sum_{<i,j>} \exp ({\rm i} \theta_{i,j} ) C_i^{\dagger} C_j +
  \sum_i W_i C_i^{\dagger}C_i ,
\end{equation}
where $C_i^{\dagger}(C_i)$ denotes a creation (annihilation)
operator  of
an electron at the site $i$.
Energies $\{ W_i \}$
are distributed independently and uniformly in the range $[-W/2, W/2]$.
By fixing the Fermi energy $E$ to be, {\it e.g.}, $E=0$ while
increasing $W$, the system is driven to be an insulator
at $W=W_c$ where $W_c$ is the critical disorder.
The Peierls phase factors $\exp ({\rm i}\theta_{i,j} )$
describe magnetic fields.
The hopping amplitude $V$ is assumed to be the
energy unit, $V=1$.
We assume a simple cubic 3D lattice for
simplicity, and all the length scales are measured
in units of its lattice constant $a$.

In the absence of magnetic fields, the AT occurs at a critical disorder
$W_c\simeq 16.5$ at the center of the band $E=0$.
The critical exponent $\nu$ has been estimated as
$\nu=1.59\pm0.03$ \cite{SO}.
Applying strong magnetic fields to the tight binding model
it has been shown that the value of the critical point as well as
the scaling curve change.  
In a magnetic field the value of $\nu$ is estimated to be
$1.43\pm 0.04$ \cite{SO}.
The exponent is not dependent of the strength of the magnetic
fields \cite{SO} and is unchanged in
random magnetic fields which can be realized by assuming a random
phase for the hopping elemnts \cite{KKO}.

Another model which describes the 3D Anderson transition
in high magnetic field is a stack of two dimensional
layers where the strong quantizing field is applied
perpendicular to the plane \cite{OOK}.
In purely two dimensional systems, the quantum Hall effect
(QHE) occurs.
The electronic states are delocalized only at the centre of
each Landau band.
In this case the critical exponent $\nu_{\rm QHE}$
is estimated to be close to 7/3 \cite{Huckestein}.
The introduction of interlayer hopping between the layers makes the
delocalized region finite in energy, and
changes the exponent to 1.45$\pm 0.15$ \cite{CD}.
This is close to the value of 1.43 given above in agreement with
universality of the AT.
Universality has also been verified for changes in the transfer integral
between the layers \cite{OOK}.

There is another interesting aspect of the stacked layer model.
If we impose Dirichlet boundary condition instead of periodic
boundary conditions, we find magnetic edge states circulating
along the perimeter of the two dimensional layer.
Stacking the layers along
the $z$-direction, we then have novel electronic states
where the electron can hop both in the $+z$ and $-z$ directions
but can rotate only in the clock-wise or anticlock-wise direction 
(which is determined by the direction of the field) in the
$x$-$y$ plane.
The edge states which compose this 
strange ^^ ^^ sheath" are critical \cite{CD,BF}.
That is to say, as we increase the linear dimension of the plane,
say $L$,
the localization length in the $z$-direction $\xi_\parallel$ also
increases in proportion to $L$ and diverges in the
thermodynamic limit.

\section{Critical Behavior}
%
%
\subsection{Anomalous diffusion and fractal dimensionality}
In the metallic regime, the electron diffuses and the mean
squared diffusion radius
$\overline{r^2(t)} \equiv 
\overline{\langle t|\vec{r}^2|t\rangle}$
is proportional to the time $t$,
\begin{equation}
\overline{r^2(t)}=2dDt ,
\label{metal_dif}
\end{equation}
where $D$ is the diffusion coefficient and the 
average is taken over disorder.
In the insulating regime where the wave function $\psi$ are localized
as $\exp(-r/\xi)$, the
squared diffusion radius saturates at \cite{deraedt}
\begin{equation}
\lim_{t\rightarrow \infty} \overline{r^2(t)} = \frac{d(d+1)}{4} \xi^2 ,
\label{ins_dif}
\end{equation}
To understand the intermediate region we use the renormalization
group, from which
\[
\overline{r^2(t)}= b^2 f((w-w_c)b^{1/\nu},t b^{-z})
\]
where $b$ is the scale factor in the renormalization group and
$z$ is the dynamical exponent.
Time $t$ is measured in units of $\hbar/V$.
From this equation we deduce the scaling form \cite{OK}
\begin{equation}
\overline{r^2(t)}= t^{2/z} F(t^{1/z\nu}(w-w_c))
\label{eqn:dif_scal}
\end{equation}
A similar relation holds for classical percolation theory \cite{Stauffer}.
On the metallic side of the transition we expect that at sufficiently
long times a linear in $t$ growth of the mean squared radius in
accordance with (\ref{metal_dif}). For this to be so we must have
$F(x)\sim x^s$ when $x\gg1$ with $s = (z-2)\nu$.
Thus on the metallic side of the transition at long times we have
\[
\overline{r^2(t)} \sim |w-w_c|^s t
\]
Since according to the Einstein relation $\sigma \sim D$ we see
that $s$ is indeed the exponent in (\ref{sexp}). 
For non-interacting electrons $z=d$ and we recover the
Wegner scaling law \cite{wegner}
\[
s = (d-2)\nu
\]
On the insulating side of the transition we expect (\ref{ins_dif}) to hold
at long times. Imposing this in (\ref{eqn:dif_scal}) leads to
\[
\overline{r^2(t)} \sim |w-w_c|^{-2\nu} \sim \xi^2
\]
confirming that $\nu$ in (\ref{eqn:dif_scal}) is indeed the exponent 
governing the divergence of the localization length.
Exactly at the critical point $w=w_c$ and we see that the 
square diffusion length grows as
\begin{equation}
\overline{r^2(t)} \sim t^{2/3} .
\end{equation}
Another quantity which can be used to investigate the dynamics of
wave function at the critical point is the return probability
\begin{equation}
C(t) \equiv \frac{1}{t}\int_{0}^{t} {\rm d}t' |\langle t'|0\rangle |^2 ,
\end{equation}
which is related to the fractal dimensionality of the wave function
$D_2$ as
\begin{equation}
C(t)\sim t^{-D_2/d} .
\end{equation}

Direct diagonalization of the 3D systems requires huge CPU power,
especially when the Hamiltonian is complex because of the applied 
magnetic fields.
Instead, we have used the equation of motion method to study the
diffusion process.
We prepare a wave packet $| 0 \rangle$ with the energy
$E$ located at the center of the system, and calculate numerically
the time evolution using $| t \rangle={\rm e}^{-{\rm i}Ht/\hbar}| 0 \rangle $.
When evaluating the factor ${\rm e}^{-{\rm i}Ht/\hbar}$,
we use the decomposition formula for exponential operators \cite{Suzuki,KO}.
In figure 1, we show the results for $\bar{r^2}(t)$
calculated for a $59\times 59\times 59$ cubic lattice.
The magnetic field is parallel to the $z$-direction and the
magnitude of the flux per unit cell is $0.1$ times the flux quantum.
The critical disorder in this case is $W_c=17.8$ \cite{HKO}.
We see clearly the $t^{2/3}$-law for $r^2(t)$, confirming that
validity of the scaling equation (\ref{eqn:dif_scal}) as well as the scaling relation
equation (\ref{eqn:wegner}).
The estimate of $D_2$ from $C(t)$ is shown in the inset of Figure 1.
We find $D_2 = 1.7$
which is significantly smaller than the spatial dimension 3,
demonstrating that the wave function at the transition is not at all
similar to a typical extended wave function.
This value is consistent with the recent estimate of $D_2$
for layered systems in high magnetic field \cite{HK}.

\subsection{Conductance distribution}
In a $d$-dimensional hyper cubic lattice
the dimension-less conductance $g$ is defined as
\begin{equation}
g=\frac{G}{e^2/h}=\frac{\sigma L^{d-2}}{e^2/h} .
\end{equation}
From the Landauer formula \cite{Landauer}, we have
\begin{equation}
g={\rm tr} t t^\dagger
\end{equation}
where $t$ is the transmission matrix.
The matrix $t$ is obtained by iteratively calculating the Green function.
\cite{Ando}.

In 3D metallic $L\times L\times L$ systems the conductance distribution 
function $P(g)$ is the normal distribution,
the mean of which $\langle g\rangle$ is proportional to the size $L$.
The variance, on the other hand, is
universal, a phenomenon which is known as universal conductance
fluctuations \cite{UCF}.
In the insulating regime $P(g)$ is log-normal.

At the Anderson transition, not only the variance
but the distribution function itself becomes universal \cite{SO,markos,Shapiro}.
The resulting distribution function is plotted in figure 2
for $L=8,10,12$ and $14$.
We plot $P(\log g)$ instead of $P(g)$ to see more clearly the
detail of the distribution function.
The histogram is for the uniform magnetic field while
the dots are for the random phase hopping model.
The critical disorder $W_c$ depends
on the strength of the field and how we break the time reversal
symmetry (i.e., uniform magnetic field or random phase hopping).
Nevertheless, $P(\log g)$ at the critical points is universal.

Once the system is away from the critical point,
the $P(g)$ begins to show size dependence.
To demonstrate this size dependence,
we plot $P(\log g)$ in vanishing field away from criticality
at $W=17.5$ in figure 3.
This value of the disorder is about 6$\%$ larger than $W_c(B=0)=16.5$.
We can see subtle but clear size dependence of $P(g)$.

Similar behavior of the conductance distribution is also observed in the layered
system in high magnetic fields,
though in this model, the system is highly anisotropic and
the form of $P_c(g)$ is different \cite{PW}.

\section{Summary and Concluding Remark}
In this paper we have discussed several features of the
Anderson transition which are related to the self-similarity of the
eigenstates at the critical point.
The fractal dimensionality of these critical eigenstates is 
almost half the original space dimension, $\approx 1.7$.
The square diffusion length $r^2(t)$ has been shown to grow as $t^{2/3}$
in the non-interacting model, irrespectively of the values of the
critical exponents.
At the transition, the distribution function of the conductance $P(g)$
becomes model and size independent.
The distribution function of $g$ close to AT had been recently obtained
experimentally \cite{PMS}, which is consistent with our results.
Such universality of the distribution is also seen in the
statistics of the energy levels at the transition \cite{Shk,ZK,OO,BSZK,KOSO}.

We have also seen that the layered system in perpendicular
magnetic fields shows interesting transport properties.
The critical behavior of this system may
still be the same as that found in the tight binding isotropic system.


In order to relate our results with experiments at finite temperature $T$,
we now discuss the $T$ dependence of the conductivity, $\sigma (T)$.
At finite temperature, the inelastic scattering time $\tau_{\rm in}$ and
the inelastic scattering length $l_{\rm in}$ are finite.
In the metallic regime, they are related by $l_{\rm in}\sim \tau_{\rm
in}^{1/2}$.
At criticality, the diffusion is anomalous diffusion and this becomes
\begin{equation}
l_{\rm in}\sim \tau_{\rm in}^{\frac{\nu}{s+2\nu}} .
\end{equation}
The effective diffusion coefficient $D_{\rm eff}$ becomes
\begin{equation}
D_{\rm eff}\sim \frac{l_{\rm in}^2}{\tau_{\rm in}}\sim \tau_{\rm
in}^{-\frac{s}{s+2\nu}} ,
\end{equation}
leading to the conductivity at finite temperature $\sigma (T)$
\begin{equation}
\sigma (T) \sim T^{\frac{s}{s+2\nu}}
\end{equation}
where we have assumed $\tau_{\rm in}\sim 1/T$.
Setting $s=\nu$ gives $\sigma\sim T^{1/3}$
which is independent of the values of the exponents $s$ and $\nu$.
In the presence of the electron-electron interaction,
the relation (\ref{eqn:wegner}) may no longer be valid and instead
\begin{equation}
s=(d-2-\theta)\nu
\end{equation}
should be used \cite{BK}.
This leads to the suggestion that the exponent of the temperature
dependence is different when the time reversal symmetry is broken.
In experiments \cite{SKUGR}
the $1/3$-power is widely observed.
This means that even in the presence of electron-electron interaction,
the relation $s=\nu$ may not be modified significantly.


\begin{figure}
\caption{
Squared diffusion length $r^2(t)$ vs. time $t$.
The solid line is the fit to
$t^{2/3}$.
Inset: double logarithmic plot of
the return probability $C(t)$ vs. $t$.
The power law $t^{-0.57}$ is a guide to the eyes.}
\end{figure}

\begin{figure}
\caption{
Distribution function of the logarithm of the conductance $g$
at the critical point.
Triangles ($\triangle$), diamonds ($\Diamond$), squares ($\square$)
and circles ($\circ$) corresponds to $L=8,10,12$ and 14
for the random phase hopping model, respectively.
The histogram is for a system in uniform magnetic fields.}
\end{figure}

\begin{figure}
\caption{
Distribution function of the logarithm of the conductance $g$
when the strength of disorder is slightly larger than the
critical value $W_c$.
Triangles ($\triangle$), diamonds ($\Diamond$), squares ($\square$)
and circles ($\circ$) again corresponds to $L=8,10,12$ and 14, respectively.}
\end{figure}

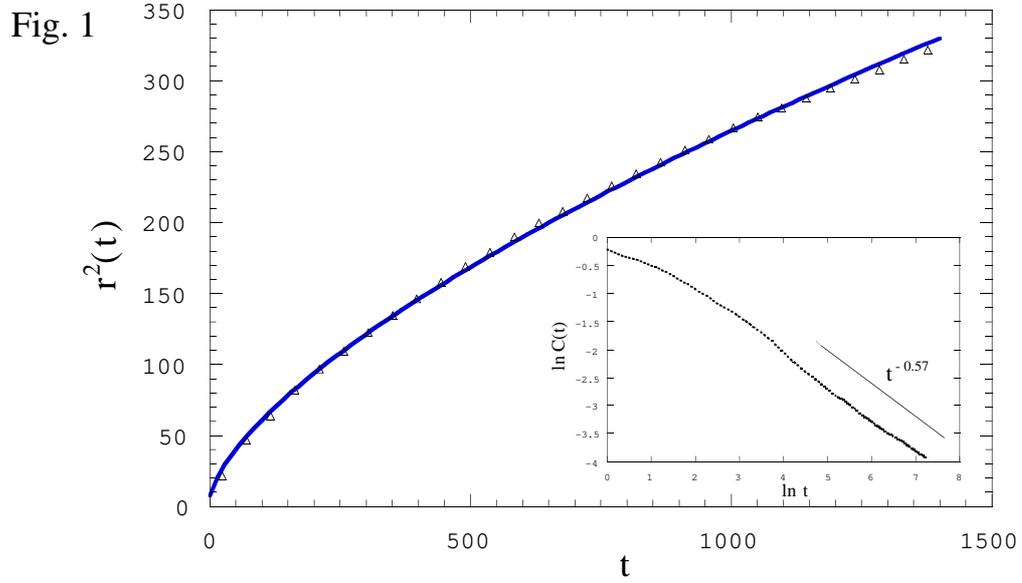
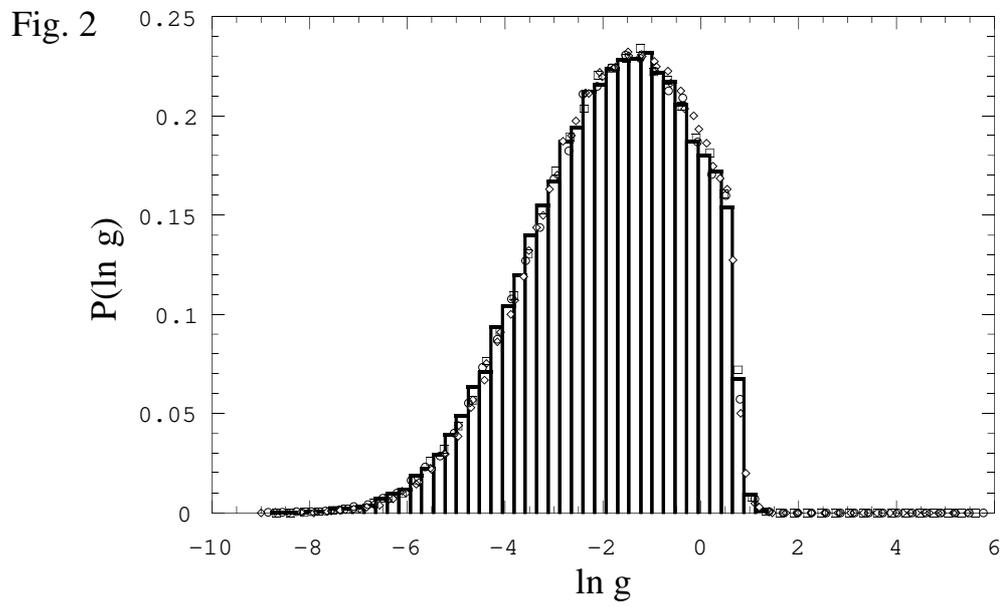
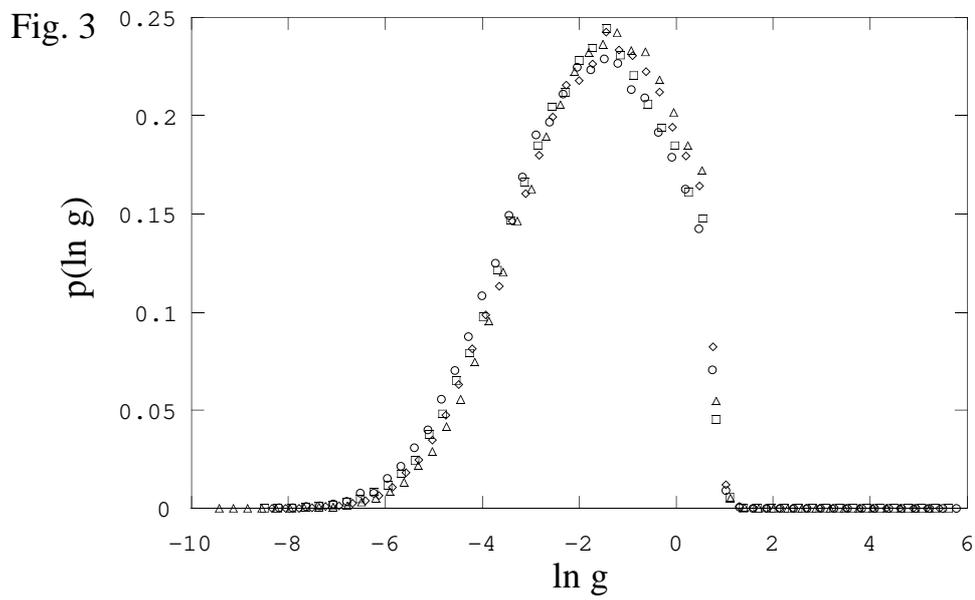
\end{document}